\documentstyle[prd,preprint,aps]{revtex}
\newcommand{\be}{\begin{equation}}
\newcommand{\ee}{\end{equation}}
\newcommand{\bea}{\begin{eqnarray}}
\newcommand{\eea}{\end{eqnarray}}

\begin{document}

\reversemarginpar
\tighten

\title{A note on quasinormal modes: A tale of two treatments} 

\author {A.J.M. Medved \thanks{E-mail:~joey\_medved@mcs.vuw.ac.nz}
and Damien Martin \thanks{E-mail:~damien\_martin@mcs.vuw.ac.nz}}

\address{
School of Mathematical and Computing Sciences\\
Victoria University of Wellington\\
PO Box 600, Wellington, New Zealand \\}

\maketitle

\begin{abstract}

There is an apparent discrepancy in the literature
with regard to the quasinormal mode frequencies
of Schwarzschild--de Sitter black holes in the
degenerate-horizon limit. 
 On the one hand, a Poschl--Teller-inspired method
predicts that the real part of the frequencies
will depend strongly on the orbital
angular momentum of the perturbation field whereas,
on the other hand,  the degenerate limit
of a monodromy-based calculation suggests
there should be no such dependence (at least, for the highly damped
modes). In the current paper, we provide a possible
resolution by critically  re-assessing  the limiting procedure
used in the monodromy analysis. 
\\
\end{abstract}

\section{Introduction}

For a long time,
a fascinating problem in gravitational physics was
 what happens to small perturbations in an otherwise
stationary black hole geometry? Fortunately, at least
the basic elements of this problem are now well understood:
Such perturbations will essentially be scattered by 
the gravitational potential and, ultimately, radiated
 away with a discrete
set of complex-valued frequencies.  Such behavior, which
is reminiscent of the last dying tones of a ringing bell, 
can be recognized   as the  quasinormal mode solutions  
associated with the black hole  spacetime. 
If nothing else, these modes are expected to be significant
in the context  of  gravitational-wave astronomy  \cite{Rev}.
  
Thanks to an intriguing (albeit conjectural) connection
between quasinormal modes and  quantum gravity,
there has been a recent surge of interest
into resolving the quasinormal mode spectra
for various black hole spacetimes. (See \cite{CLrev,newpad}
for  summarized accounts  of recent work and the relevant references.)
In a sense, such a connection is surprising, given
that quasinormal modes represent a purely classical
consequence of the black hole gravitational field.
Nonetheless, as first  advocated by Hod \cite{Hod},
there is reason to believe that quasinormal modes
can be used to fix the level spacing of
the  black hole area spectrum. (An evenly spaced area
spectrum follows from  arguments, as made initially  
by Bekenstein \cite{Bek},
that the horizon area of a black hole is an adiabatic invariant.)
Moreover, this rationale can even  be extended into the realm
of loop quantum gravity, where it has been suggested by Dreyer \cite{Dre}
that similar considerations  can be utilized to fix, unambiguously,
the elusive ``Immirzi parameter'' \cite{Imm}.

In the above  circle of ideas, a pivotal  role is
played by  the  real part of the quasinormal mode frequency 
when the modes are highly damped ({\it i.e.}, when the
magnitude of the imaginary part  is very large).
It is generally believed (or perhaps hoped)
 that, for a given black hole spacetime,
 this real part will asymptote to
a fixed value, independently of the details of the perturbation.
In  light of this comment, it is instructive to consider a specific
model;  let us therefore  call upon  the Schwarzschild mode
spectrum for illustrative purposes. In this case,
one obtains (with appropriately chosen boundary conditions)
 the following set of frequencies 
for either scalar or axial gravitational
perturbations \cite{Nol,And,Motl,MN,Neitzke}:
\be
k_{qnm}=\kappa\left[i\left(n+{1\over 2}\right)+{1\over 2\pi}\ln 3\right]
+{\cal O}[n^{-1/2}] \quad\quad n=0,1,2,...\;.
\label{0}
\ee
Note that $\kappa$ is the (Schwarzschild) surface gravity~\footnote{Here
and throughout, all fundamental constants have been set to unity.} 
and this expression  becomes increasingly accurate as $n\rightarrow\infty$.

This (asymptotically valid) spectral form has long been known by 
numerical means ({\it e.g.}, \cite{Nol}) and has even, quite recently,
 been confirmed by analytical treatments.
In \cite{MN,Neitzke}, for instance, the authors have  invoked
a method that is based on calculating the monodromy
 of the perturbed field when the radial coordinate has been
  analytically continued to
the complex plane.  In any event, 
 the above spectrum  substantiates that the real
part of the frequencies does indeed asymptote to a constant value,
independent of any details about the perturbation
field itself.~\footnote{Actually, the type
of  perturbation  field does play a role, as
the above form applies, strictly speaking,  to only
 scalar and axial gravitational
perturbations.  Nonetheless, it could  be argued
that fundamental considerations should be restricted
to a certain class of gravitational perturbations, so this is 
generally not  a concern.}
 As a further point of interest, 
 it is a generic feature of
black hole spacetimes that the asymptotic spacing (between
the imaginary levels) goes precisely  as the surface gravity 
\cite{MVW,PAD}.

Let us re-emphasize that the real part  of the (highly damped)
quasinormal frequencies is a critical ingredient in the
proposals made by both Hod \cite{Hod} and Dreyer \cite{Dre}.
Moreover, the stance of these authors  is
that this  real portion   represents
a fundamental (transition) frequency 
associated with the black hole horizon.
Hence, if their arguments are to hold up, it is essential
that this real part is strictly  characterized by
the horizon geometry and, hence,
 not overly sensitive to specifics of the perturbation
field. 
It does, however, remain somewhat controversial as
to whether or not this ``non-sensitivity'' is indeed
a universal feature of black hole spacetimes.

For instance, a  viable counterexample is provided 
by  Schwarzschild--de Sitter space when the
black hole and cosmological horizon are closely
``squeezed'' together.~\footnote{We often refer
to the degenerate  (or very nearly degenerate) Schwarzschild--de Sitter model
as being in the ``squeezed-horizon'' limit \cite{squeeze}.  Such terminology 
is meant to distinguish 
this scenario from one in which the observer is
exterior to a pair of (nearly) coincident horizons; that is,
exterior to a  black hole at (or near) extremality.  The two
scenarios are, of course, operationally distinct.}
To elaborate, a recent study \cite{CL} ---
 based on identifying the relevant scattering potential 
with that of the Poschl--Teller model \cite{PT,FM} 
%(for
%which the quasinormal mode spectrum is  known exactly \cite{FM})
 ---
found a quasinormal spectrum that  
depends strongly on the orbital angular momentum ($\ell$)  of
the perturbation field. Moreover, in the degenerate limit, 
the real part of the frequency, for {\em any} mode, goes almost linearly
with $\ell$. (In fact, it has since been demonstrated 
 that this behavior  is a generic  feature
of squeezed-horizon spacetimes \cite{squeeze}.)
However,   this is not yet the full story.  In 
a more  recent paper \cite{CBA},
the quasinormal
mode spectrum was calculated  for  non-degenerate 
Schwarzschild--de Sitter space by way of the monodromy method
\cite{MN,Neitzke}.   When this form of the  spectrum
is then subjected to the horizon-degeneracy limit, as was done explicitly
in \cite{CBA},  there  is absolutely no $\ell$
dependence in evidence. Hence, what  we have is two quite
conflicting predictions for precisely the same 
model.

The purpose of the current paper is to provide a possible
resolution for this rather disturbing discrepancy. Our
basic point of view is that the Poschl--Teller-inspired calculation
is, given its elegant simplicity, most likely correct.\footnote{Moreover,
the Poschl--Teller spectral
 form  for nearly degenerate Schwarzschild--de Sitter
space has also been substantiated semi-analytically 
(particularly for
large values of $\ell$) by  Moss and Norman \cite{Mono} using 
 continued-fraction
techniques \cite{Leav}, as well as by  Zhidenko \cite{Zhi} and
Konoplya \cite{Kon3} using  WKB methodology \cite{SW,IW,Kon2}.
For other recent studies on Schwarzschild--de Sitter quasinormal modes,
 see \cite{Sun,Mol,Brink,YF}.} 
 Meanwhile, the monodromy-based  calculation, although perfectly
valid in the non-degenerate regime, can not necessarily
be extrapolated up to the point of horizon coincidence. 
The remainder of the paper  contains the
analysis in support of our argument (Section II), followed
by a pertinent  discussion (Section III). Note that
we skip over most of  the  details  of these interesting methodologies
--- ``Poschl--Teller'' \cite{CL,squeeze} and ``monodromy''
\cite{MN,Neitzke,CBA} --- and refer the reader to the
cited works.

\section{Analysis}

Before  getting to the crux of the matter, let
us introduce some necessary  formalism.
The metric for a Schwarzschild--de Sitter
spacetime can be expressed (in a static coordinate
gauge) as follows:
 \be
d s^2 = -  f(r) d t^2 
+ f^{-1}(r) d r^2 + r^2 d \Omega^2 \; ,
\label{1}
\ee
where 
\be
f(r)=1-{2M\over r}-{r^2\over a^2}\;,
\label{2}
\ee
 with $M$ representing the black hole mass and $a$ denoting
the de Sitter curvature radius (alternatively, $\Lambda=3/a^2$
is the positive cosmological constant). 

 The function $f(r)$ has
three zeroes, two of which locate the black hole
horizon, $r_b$, and the cosmological horizon, $r_c$,
with $r_b\leq r \leq r_c$  indicating the ``observable''
portion of the
spacetime.
[The third zero locates a fictitious negative horizon,
$r_0=-\left(r_b+r_c\right)$.] For future reference,
it is useful to take note of the following relations:
\be
a^2=r_b^2+r_br_c+r_c^2\;,
\label{2.5}
\ee
\be
2Ma^2=r_br_c\left(r_b+r_c\right)\;.
\label{2.75}
\ee

Each horizon is endowed with a surface gravity, which
can be evaluated via standard methods 
(essentially, one half of the derivative of $f(r)$
evaluated at the appropriate horizon \cite{gibhaw})
to yield
\be
\kappa_b={\left(r_c-r_b\right)\left( 2r_b +r_c\right)
\over 2a^2 r_b} \;,
\label{3}
\ee
\be
\kappa_c={\left(r_c-r_b\right)\left( 2r_c +r_r\right)
\over 2a^2 r_c} \;,
\label{4}
\ee
for the black hole and cosmological horizon respectively.
Note that these quantities are, by definition, positive definite.

It is often convenient --- especially, in the context
of quasinormal modes --- to introduce a (generalized) ``tortoise
coordinate''; that is, $dx=dr/f(r)$ or
\be
x=\int {dr\over f(r)}\;.
\label{5}
\ee
Substituting in equation (\ref{2}) and integrating, one obtains
\be
x={1\over 2\kappa_b}\ln\left[{r\over r_b}-1\right]
-{1\over 2\kappa_c}\ln\left[1-{r\over r_c}\right]
+{1\over 2}\left({1\over \kappa_c}- {1\over\kappa_b}\right)
\ln\left[{r\over r_b+r_c}+1\right] \;.
\label{6}
\ee
Keep in mind that 
$r\in\left(r_b,r_c\right)$
maps into the region $x \in\left(-\infty,+\infty\right)$.

Let us now focus specifically on the quasinormal mode
problem, which entails the study of how small (massless) perturbations
 of the background spacetime are scattered by
the gravitational potential. As has been well documented,
one can formally cast this picture into a one-dimensional,
Schrodinger-like scattering situation \cite{RW,Rev}. 
Generically speaking, one obtains an equation that conforms to
\be
{d^2\over d x^2} \psi - V[r(x)] \, \psi = - k^2 \psi\;,
\label{7}
\ee
where $\psi=\psi\left[r(x)\right]$  describes
the radial behavior
of the  perturbation field, $V[r(x)]$  is a model-dependent
``scattering potential'',  and $k$ is
the frequency (that is, $e^{\pm ikt}$ is
the time dependence of the perturbation; with $e^{+ikt}$ then chosen
 to ensure that the imaginary part of $k$ is
positive for an exponentially decaying solution).
Note that the potential generally
depends on both the spin, $j$, and the 
orbital angular momentum, $\ell$, of the
perturbed field.

For a Schwarzschild--de Sitter spacetime, 
in particular, it has been  shown that the
scattering potential takes on the form \cite{carlem}
\be
V[r]=f(r)\left[{\ell\left(\ell+1\right)\over r^2}
+\left(1-j^2\right)\left({2M\over r^3}-{2\over a^2}
\right)\right]\;,
\label{8}
\ee
at least for  scalar ($j=0$) and axial gravitational
($j=2$) perturbations. (Generalizations to other
cases are indeed possible \cite{carlem} but, for simplicity, will not be
considered here.)

As a brief but  important aside, let us point out that
 resolving the quasinormal mode problem requires
suitably chosen initial conditions. Normally, one 
imposes ``radiation boundary conditions'' such 
that $\psi(x)\propto e^{\pm ikx}$ as $x\rightarrow\mp\infty$;
that is, an ingoing (outgoing) plane wave at the
inner (outer)  boundary.  It is also necessary,
in the case of dual-horizon scenarios, to make a decision
as to where the scattering takes place. In the case of 
{\em black hole} scattering, the  ``incident'' wave should
be coming in from the outer boundary ({\it i.e.}, positive
infinity in $x$ coordinates)   whereas, for {\em cosmological  
horizon} scattering, the incident wave should be coming out
from the inner boundary. (For a more quantitative description,
see \cite{newpad}.) In the (current) case
of nearly coincident horizons, however, this distinction
becomes effectively irrelevant.

Since our current interest is specifically with the nearly degenerate
(or ``squeezed'') horizon scenario, we will now  
re-express the above formalism as appropriate for this regime.
First, let us put the notion of nearly coincident horizons   into
quantitative terms. This can be accomplished with
the introduction of  the following ``squeezing
parameter'':
\be
\Delta\equiv {\left(r_c-r_b\right)\over r_b} << 1\;.
\label{9}
\ee
Given the extent of the relevant manifold
($r_b\leq r\leq r_c\;$), it immediately
follows that, up to corrections of the relative order $\Delta\;$,
$r_b\sim r \sim r_c\;$ and $\kappa_b\sim\kappa_c\;$. 
(For the duration, $\sim$ will always  be used to signify corrections
of this order.)

In view of the above, the metric function (\ref{2}), surface gravities
(\ref{3},\ref{4}), tortoise coordinate (\ref{6}) and
scattering potential (\ref{8}) will  now  simplify
as follows:
\be
f(r)\sim 2\kappa_b(r-r_b)\;,
\label{10}
\ee
\be
\kappa_b\sim\kappa_c\sim{\left(r_c-r_b\right)
\over 2 r_{b}^2} \;,
\label{11}
\ee
\be
x\sim{1\over 2\kappa_b}\left(\ln\left[{r\over r_b}-1\right]
-\ln\left[1-{r\over r_c}\right]
+\left({\kappa_b-\kappa_c\over \kappa_b}\right)
\ln\left[{r\over r_b+r_c}+1\right]\right) \;,
\label{12}
\ee
\be
V[r]\sim f(r){\ell\left(\ell+1\right)\over r^2}\;.
\label{13}
\ee
Take particular note of the scattering potential;
it depends,  up to corrections that vanish as $\Delta\rightarrow 0$,
on the orbital angular momentum  but {\em not}  on
the spin of the perturbation.

Given this simplified  form of the  potential, it turns out  
that one can directly extract  the quasinormal mode frequencies
from equation (\ref{7}).
More to the point, as shown in \cite{CL} 
(and later generalized in \cite{squeeze}), the potential 
$V[x]$  takes on a Poschl--Teller form \cite{PT},
for which the quasinormal mode spectrum is known exactly \cite{FM}.
Following this reasoning, one eventually obtains \cite{CL}
\be
k_{qnm}=\kappa_b \left[i\left(n+{1\over 2}\right)
+\sqrt{\ell(\ell+1)-{1\over 4}}
+{\cal O}[\Delta]\right]
\quad\quad n=0,1,2,...\;. 
\label{14}
\ee

Next, we would  like to know what the 
monodromy method \cite{MN,Neitzke} can say
 about the quasinormal modes for Schwarzschild--de Sitter space
when the horizons are closely squeezed.
In fact, this  monodromy
 calculation has already been carried out in \cite{CBA}, although
under the presumption of  non-degenerate horizons. 
Still,  the authors of \cite{CBA} considered (near)  horizon degeneracy
as a limiting case of the general analysis.
 It should, however, be emphasized that the appropriate limit
was taken   only after the calculation of the mode spectrum was completed.
We will now proceed to argue that this particular limiting
procedure may  not be  technically correct

One of the underlying premises of  the monodromy method \cite{MN,Neitzke} is 
the viability of an analytic continuation  of $r$ (and, by implication, $x$) 
into the complex-valued plane. This continuation allows
the real part of $r$ to enter the ``unobserved'' region
($r<r_b$) including all the way to the singularity
at $r=0$.  However, as pointed out in \cite{newpad},
it is essential to the program  that the geometry
(specifically, $f(r)$ and related quantities)
is {\em first} defined in the observable region ($r\geq r_b$)
and {\em then}  analytically continued. That is to say,
the ``physical'' geometry of the black hole interior
[{\it i.e.}, $f(r)$ as defined for $r<r_b$  {\it sans} any continuation] 
must {\it not} be allowed to  enter into the quasinormal mode problem.
This is because, in the original definition of
the problem, one sets up strict  boundary conditions
at the black hole horizon and the cosmological horizon,~\footnote{For
as asymptotically flat spacetime, the boundary conditions
are rather fixed at the black hole horizon and spatial infinity.}
 thus rendering  the black hole interior  (and the cosmological horizon 
exterior)
 as being operationally irrelevant.

In view of the above discussion, it becomes clear that
a ``squeezed observer'' would find it most appropriate to
 analytically continue
 the geometry  described 
by equations (\ref{10}-\ref{13}). (Admittedly, this is somewhat naive 
--- see Section III for an elaboration.)
Now, considering  that the ``focal point''
 of the
monodromy calculation is at (or rather near) the singularity, $r=0$,
let us see what happens to these expressions near the
spatial origin.  Firstly, defining a ``shifted tortoise
coordinate'' or $z=x+constant\;$ such that $z(r=0)=0$ \cite{MN},
we are able to deduce 
[after  expanding equation (\ref{12})  near $r=0$ and
then incorporating equation (\ref{11})] 
\be
z\sim -{3\over 2} r + {\cal O}[r^2]\;.
\label{15}
\ee

Next, let us consider the potential [{\it cf}, equations (\ref{10},\ref{13})]
  as $r$ or $z$ goes to zero. This is simply
\bea
V[z(r)]&\sim& -2{r_b\kappa_b\ell\left(\ell+1\right)\over r^2}
+{\cal O}[r^{-1}]\nonumber \\
&\sim&  -{9\over 2}{r_b\kappa_b\ell\left(\ell+1\right)\over z^2}
+{\cal O}[z^{-1}]\;.
\label{16}
\eea
Conveniently, this is  the same form of
near-the-origin potential ($V\propto z^{-2}$) as obtained
 for the non-degenerate Schwarzschild--de Sitter
case \cite{CBA}. Hence, the rest of the monodromy-based
 calculation can proceed, 
in the same manner as \cite{CBA},
but with the necessary  substitution
\bea
\nu &=& {1\over 2}\sqrt{1+z^2V[z\approx 0]} \nonumber \\
&=& {1\over 2}\sqrt{1-{9\over2}r_b\kappa_b\ell\left(\ell+1\right)},
\label{17}
\eea
where $\nu$ is the ``Bessel-function index'' as explicitly
defined in equations (12-13) of \cite{CBA}.

Referring the reader  again to \cite{CBA} (also see \cite{MN,Neitzke}), 
let us quote the quasinormal mode spectrum as obtained
via  the prescribed procedure: $e^{2\pi k/\kappa_b}=
-\left(1+2\cos 2\pi\nu\right)$ or
\be
k_{qnm}=\kappa_b \left[i\left(n+{1\over 2}\right)
\pm{1\over 2\pi}\ln \left| 1+2\cos 2\pi\nu \right| \right]   \quad\quad
n=0,1,2,...
\;,
\label{18}
\ee
which can  be expected to be valid up to corrections of
the order $n^{-1/2}$ (in general) and  the relative order $\Delta$ (in
our specific case).
Substituting in  equation (\ref{17}) and expanding under
the  assumption that ${9\over 2}r_b\kappa_b\ell\left(\ell+1\right)$
is somewhat less than unity,~\footnote{Since $r_b\kappa_b\approx \Delta$ 
and $\Delta<<1$,
this assumption must be true as long as 
 $\ell$ is not too large.}
we then have
\be
k_{qnm}=\kappa_b \left[i\left(n+{1\over 2}\right)
\pm {91\over 32}\pi r_b^{2}\kappa_b^2\ell^2\left(\ell+1\right)^2 \right]   
\quad\quad {\rm as} \quad n\rightarrow\infty  \quad {\rm and}
\quad \Delta\rightarrow 0\;.
\label{19}
\ee

\section{Discussion}

Some commentary is, of course, in order.
First of all, it is interesting to compare this last (monodromy)
calculation with
the spectrum  obtained from the ``Poschl--Teller method'' 
\cite{CL}  [as depicted in equation (\ref{14})].
We immediately see that the asymptotic spacing 
between the imaginary levels is correctly
realized; an outcome which is, in  itself,  by no means trivial.
On the other hand, the real part of the (asymptotic) frequency is
somewhat different; in particular, take note of the
quartic  versus linear dependence of $\ell$.
It is, however, possible 
 to explain this discrepancy as follows (also see the Addendum):
 
It is not quite  clear
that equation (\ref{13}) is truly the correct form of
 the potential that  should  extrapolated down to
$r=0$.  More to the point,  one can always  add additional terms 
to the potential  that do {\em not} significantly 
alter its structure in the observable region but which
would, nevertheless, dominate 
at the origin and, thus,  make a significant contribution to the monodromy.
That is to say, more care is needed in addressing the
contributions  from such  ``neglected'' terms.  To understand
the type of analysis that might be required,
one can turn to a close cousin of the monodromy
calculation; namely, the WKB method of quantum mechanics.
As discussed in \cite{NHow}, 
the Stokes and anti-Stokes lines (with three of each emanating
from the $r=0$ singularity) need to be addressed.
We will defer the fully  rigorous treatment of the monodromy calculation
 until  a later time.
It does, however,
%the monodromy-based calculation of 
%\cite{CBA} can be extended, in a straightforward manner,
% to the (nearly) degenerate-horizon scenario.
%More specifically, the monodromy calculation
%follows a  contour of integration which, arguably,
%becomes poorly defined in the limit
%that the two horizons coincide. (For instance,
%a portion of the contour follows a ``large semi-circle'',
%$|r|\rightarrow |r_c|\;$, which is presumed to 
%be reasonably far from the black hole horizon.
%This is obviously not the case under the current circumstances.)
%How to address these analytical caveats is, at present,
%an unresolved problem.
%Nonetheless, it remains clear that, in the squeezed-horizon limit,
% the  real part
%of the frequency will depend strongly on the orbital angular
%momentum
%[$\ell$  being prominent in the only surviving term in the potential;
%{\it cf}, equation (\ref{13})]. 
 seem 
feasible that such a treatment  could still
 give back equation (\ref{14}) [that is, it does not
appear to be the naively simple matter of just
summing up the various contributions].
Such speculation aside, the  point we are really trying to make is that, 
 as $\Delta\rightarrow 0$,
 the $\ell$-dependent term clearly dominates the potential  so that, regardless
of any neglected contributions, this $\ell$-dependence  can not 
just be  
blatantly ignored.\footnote{Alternatively, one might consider
the quasinormal mode problem strictly in the observable region
(as advocated by Born-approximation-inspired methods \cite{MVW,PAD}),
where the $\ell$-dependence is so evidently prominent.}

Secondly, it is of interest to see how things match up
when equation (\ref{19}) is compared with the  spectral form
predicted by the (``non-degenerate'') monodromy calculation
of \cite{CBA}.  To reiterate, these authors obtain equation (\ref{18})
but with a different expression for the index $\nu$.
To be precise, they have found
\be
\nu={1\over 2}\sqrt{1+(4M\beta)^2\left(j^2-1\right)}\;,
\label{20}
\ee
where $\beta$ is a parameter that depends strictly
on $r_b$, $r_c$,  $\kappa_b$ and $\kappa_c$.
The philosophy of \cite{CBA} was  that the (nearly) degenerate
limit could then  be dealt with by directly extrapolating
 this  non-degenerate form for  the spectrum.
Following this approach, one observes  that
the quasinormal frequencies of  (nearly) degenerate Schwarzschild--de Sitter
space  can certainly depend
on the spin, $j$,  but definitely  {\em not} on the orbital
angular momentum, $\ell$. How can we make sense of this
discrepancy in the $\ell$ dependence of the frequencies?
Well, although  a
reasonable argument could be made  both ways --- that is to say, it is not
particularly  clear what is the rigorously  correct limiting procedure 
--- strong support for an $\ell$-dependent  spectrum
 follows from the Poschl--Teller calculation.
Significantly, this methodology --- which is unambiguous and
exact in the squeezed-horizon limit ---  implies that
the real part of the  frequency can   depend
 {\it only}  on  $\ell$ and $\kappa_b$ when the
horizons coincide.

Let us now make  a couple of   general observations.
Firstly, we can see from the subtleties of
the current case that other quasinormal mode calculations 
that go on to apply extremal (or near-extremal)
limits  should probably be viewed with a healthy
dose of scepticism. Which is to say, the precise moment
(in the calculation) that such a limit can safely  be enforced
would appear to be a  subject worthy  of  debate.  
Secondly, let us re-emphasize 
that a strongly $\ell$-dependent   quasinormal mode spectrum
(particularly, in the real part of the highly damped frequencies)  
should  cast significant 
doubt  on the  status  of  quasinormal modes in quantum
gravity.  (Let us recall that the real part of the frequency
can be used, conjecturally  speaking, to fix
the spacing of the black hole area spectrum \cite{Hod}
and, by implication,  
the Immirzi parameter of loop quantum gravity \cite{Dre}.)
It is not clear, at least to the current authors, 
how the fundamental quantum theory  can be sensitive
to the details of the perturbation field or, alternatively,
why would only certain classes of horizons  be
susceptible to quantization?

Finally, it is worthwhile to point out another discrepancy,
in the literature, which centers around the Schwarzschild--de Sitter
black hole spacetime.  With emphasis on a scenario of non-degenerate
horizons, it has been argued (by the current authors and M. Visser \cite{MVW}
--- also see \cite{Sun})
that both the cosmological and the black hole horizon
will ``contribute'' to the quasinormal mode spectrum. That is,
we  anticipate one set of modes that goes (asymptotically)
as $in\kappa_b$  and another set that goes as $in\kappa_c$.
Our argument is essentially that, from the perspective of
an observer in this spacetime, there would be no reason
to give one horizon a preferred status over the other.
Hence, for a {\em complete} calculation, it would be necessary to
 deal with two separate
scattering problems which can be distinguished
by the choice of initial conditions (see the related discussion
in Section II).
Nonetheless, a contrary opinion has  been 
expressed in \cite{newpad}, where it is claimed that the cosmological horizon
scattering conditions would not be physically relevant.
The current authors, however, can see no  convincing
reason why one horizon should be singled out {\it a priori}.
Unfortunately, any study which focuses on the case
of horizon degeneracy can offer nothing substantial
to this particular argument. Hence, we have nothing
to add at this time  but hope
to readdress the matter, by more rigorous means, in 
the near future.

\section*{Addendum}

 It has been brought to our attention,
in a response to the first archival version of
the paper \cite{X1,X2},
that the  monodromy calculation used in \cite{CBA} 
is not quite correct.
The main sticking point seems to be that, upon
continuation to the complex plane, 
the generalized tortoise coordinate
does {\em not}  go to infinity as $r$ goes to
$r_c$; thus invalidating part of  the matching procedure
used in \cite{CBA}. Given this breakdown, it is
even less surprising that our  monodromy
result (\ref{19}) fails
to  reproduce the Poschl--Teller calculation (\ref{14}).
It should also be emphasized that the invalidation
of the original (Schwarzschild--de Sitter) monodromy analysis
in no way undermines any of our prior discussion.
To elaborate: in \cite{CBA}, the near-the-origin scattering  potential
depends on $j$ but not $\ell$, whereas our analogue
depends on $\ell$ but not $j$.  So, even if equation (\ref{18})
[that is, the spectrum predicted by the original
monodromy calculation] is wrong,  our ``corrected form''
must necessarily differ from that of \cite{CBA} in a substantial
way. Namely, $\ell$ dependence versus no $\ell$ dependence.

\section*{Acknowledgments} 
\par
The authors thank M. Visser for guidance.  The
authors also thank K. Castello-Branco, R. Konoplya and
J.P. Norman for helpful comments with regard
to the first version of the paper. Research is supported by
the Marsden Fund (c/o the New Zealand Royal Society) 
and  the University Research Fund (c/o Victoria University).
\par


\vspace*{20pt}

\begin{thebibliography} {99}

\bibitem{Rev}  See, for comprehensive reviews and important references: \\
H.-P. Nollert,
``Quasinormal modes: the characteristic sound of black holes 
and neutron stars'',
 Class. Quant. Grav. {\bf 16}, R159 (1999);
%%This citation has NO SPIRES listing??  
\\
K.D. Kokkotas and B.G. Schmidt, 
``Quasi-Normal modes of black holes and stars'',
Living Rev. Rel. {\bf 2}, 2 (1999) [arXiv:gr-qc/9909058].
%%CITATION = GR-QC 9909058;%%


\bibitem{CLrev} 
V. Cardoso, J.P.S Lemos and S. Yoshida, 
``Quasinormal modes of Schwarzschild black holes in four
and higher dimensions'', arXiv:gr-qc/0309112 (2003).
%%CITATION = GR-QC 0309112;%%


\bibitem{newpad} T.R. Choudhury and T. Padmanabhan,
``Quasi normal modes in Schwarzschild-DeSitter spacetime:
A simple derivation of the level spacing of the frequencies'',
arXiv:gr-qc/0311064 (2003).
%%CITATION = GR-QC 0311064;%%


\bibitem{Hod} 
S.~Hod,
``Bohr's correspondence principle and the 
area spectrum of quantum black  holes'',
Phys.\ Rev.\ Lett.\  {\bf 81},  4293 (1998)
[arXiv:gr-qc/9812002].
%%CITATION = GR-QC 9812002;%%

\bibitem{Bek} 
J.D.~Bekenstein,
``The Quantum Mass Spectrum Of The Kerr Black Hole'',
Lett.\ Nuovo Cim.\  {\bf 11}, 467 (1974).
%%CITATION = NCLTA,11,467;%%

\bibitem{Dre} 
O.~Dreyer,
``Quasinormal modes, the area spectrum, and black hole entropy'',
Phys.\ Rev.\ Lett.\  {\bf 90}, 081301 (2003)
[arXiv:gr-qc/0211076].
%%CITATION = GR-QC 0211076;%%

 
\bibitem{Imm} G. Immirzi, ``Quantum Gravity and Regge Calculus'',
 Nucl. Phys. Suppl. {\bf 57},
65 (1997) [arXiv:gr-qc/9701052].
%%CITATION = GR-QC 9701052;%%




\bibitem{Nol} 
H.-P. Nollert,
``Quasinormal modes of Schwarzschild black holes: the determination
of quasinormal frequencies with very large imaginary parts'',
 Phys. Rev. D {\bf 47}, 5253 (1993).
%%This citation has NO SPIRES listing??


\bibitem{And}  
N. Andersson, ``On the asymptotic distribution of
quasinormal-mode frequencies for Schwarzschild black holes'',
Class. Quant. Grav. {\bf 10}, L61 (1993).
%%This citation has NO SPIRES listing??


\bibitem{Motl} 
L.~Motl,
``An analytical computation of asymptotic 
Schwarzschild quasinormal  frequencies'',
Adv.\ Theor.\ Math.\ Phys.\  {\bf 6}, 1135 (2003) 
[arXiv:gr-qc/0212096].
%%CITATION = GR-QC 0212096;%%


\bibitem{MN}
L.~Motl and A.~Neitzke,
``Asymptotic black hole quasinormal frequencies'',
Adv.\ Theor.\ Math.\ Phys.\  {\bf 7}, 307 (2003) 
[arXiv:hep-th/0301173].
%%CITATION = HEP-TH 0301173;%%

\bibitem{Neitzke}
A. Neitzke, ``Greybody factors at large imaginary frequencies'',
arXiv:hep-th/0304080 (2003).
%%CITATION = HEP-TH 0304080;%%



\bibitem{MVW}  A.J.M. Medved, D. Martin and M. Visser,
``Dirty black holes: Quasinormal modes'', arXiv:gr-qc/0310009
(2003).
%%CITATION = GR-QC 0310009;%%


\bibitem{PAD} T. Padmanabhan, ``Quasi normal modes: A simple
derivation of the level spacing of the frequencies'',
arXiv: gr-qc/0310027 (2003).
%%CITATION = GR-QC 0310027;%%


\bibitem{squeeze} A.J.M. Medved, D. Martin and M. Visser,
``Dirty black holes: Quasinormal modes for ``squeezed horizons'',
arXiv:gr-qc/0310097 (2003).

\bibitem{CL} V. Cardoso and J.P.S. Lemos, ``Quasinormal modes of the
near extremal Schwarzschild-de Sitter black hole'',
Phys. Rev. D {\bf 67}, 084020 (2003) [arXiv:gr-qc/0301078].
%%CITATION = GR-QC 0301078;%%



\bibitem{PT} G. Poshl and E. Teller, Z. Phys. {\bf 83}, 143
(1933).
%%This citation is pre-SPIRES


\bibitem{FM} V. Ferrari and B. Mashhoon, Phys. Rev. D {\bf 30},
295 (1984).
%%This citation has NO SPIRES listing??


\bibitem{CBA} K.H.C. Castello-Branco and E. Abdalla,
``Analytic determination of the asymptotic quasi-normal mode spectrum
of Schwarzschild-de Sitter black holes'', arXiv:gr-qc/0309090 (2003).
%%CITATION = GR-QC 0309090;%% 


\bibitem{Mono} I.G. Moss and J.P. Norman, ``Gravitational quasinormal
modes for anti-de Sitter black holes'', Class. Quant. Grav. {\bf 19},
2323 (2002) [arXiv:gr-qc/0201016].
%%CITATION = GR-QC 0201016;%%

\bibitem{Leav} E.W. Leaver, ``An analytic representation for
the quasi-normal modes of Kerr black holes'', Proc. Roy. Soc.
Lond. A {\bf 402}, 285 (1985).
%%CITATION = PRSLA,A402,285;%%

\bibitem{Zhi} A. Zhidenko, ``Quasi-normal modes of Schwarzschild-de
Sitter black holes'', arXiv:gr-qc/0307012 (2003).
%%CITATION = GR-QC 0307012;%%


\bibitem{Kon3} R.A. Konoplya, ``Gravitational quasinormal radiation
of higher-dimensional black holes'', arXiv:hep-th/0309030 (2003).
%%CITATION = HEP-TH 0309030;%%

\bibitem{SW} B.F. Schutz and C.M. Will, ``Black hole normal modes:
A semianalytic approach'', Astrophys. J. {\bf 291}, L33 (1985).
%%This citation has no spires listing 


\bibitem{IW} S. Iyer and C.M. Will, ``Black hole normal modes: A WKB
approach'', Phys. Rev. D {\bf  35}, 3621 (1987).
%%CITATION = PHRVA,D35,3621;%%



\bibitem{Kon2} R.A. Konoplya, ``Quasinormal behavior of the D-dimensional
Schwarzschild black hole and higher order WKB approach'',
Phys. Rev. D {\bf 68}, 024018 (2003) [arXiv:gr-qc/0303052].
%%CITATION = GR-QC 0303052;%% 


\bibitem{Sun}
V. Suneeta, 
``Quasinormal modes for the SdS black hole: 
an analytical approximation scheme'', 
Phys. Rev. D {\bf 68}, 024020 (2003)
[arXiv:gr-qc/0303114].
%%CITATION = GR-QC 0303114;%%


\bibitem{Mol} C. Molina, ``Quasinormal modes of d-dimensional spherical
black holes with a  near extreme cosmological constant'',
Phys. Rev. D {\bf 68}, 064007 (2003) [arXiv:gr-qc/0304053].
%%CITATION = GR-QC 0304053;%%

\bibitem{Brink} A. Maassen van den Brink, ``Approach to the extremal
limit of the Schwarzschild-de Sitter black hole'', Phys. Rev. D
{\bf 68}, 044024 (2003) [arXiv:gr-qc/0304092].
%%CITATION = GR-QC 0304092;%%


\bibitem{YF} S. Yoshida and T. Futamase, ``Numerical Analysis
of quasinormal modes in nearly extremal Schwarzschild-de Sitter
spacetimes'', arXiv:gr-qc/0308077 (2003).
%%CITATION = GR-QC 0308077;%%








\bibitem{gibhaw} G. Gibbons and S.W. Hawking, ``Action integrals and
partition functions in quantum gravity'', Phys. Rev. D {\bf 15},
2752 (1977); 
%%CITATION = PHRVA,D15,2752;%%
``Cosmological event horizons, thermodynamics and
particle creation'', Phys. Rev. D {\bf 15}, 2738 (1977).
%%CITATION = PHRVA,D15,2738;%%

\bibitem{RW} 
T. Regge and J.A. Wheeler, 
``Stability of a Schwarzschild singularity'',
Phys. Rev. {\bf 108}, 1063 (1957).
%%This citation is pre-SPIRES




\bibitem{carlem} V. Cardoso and J.P.S. Lemos, ``Quasi-normal modes
of Schwarzschild anti-de Sitter black holes: Electromagnetic
and gravitational perturbations'',  Phys. Rev. D {\bf 64}, 084017
(2001) [arXiv:gr-qc/0105103].
%%CITATION = GR-QC 0105103;%%

\bibitem{NHow} N. Andersson and C.J. Howls, ``The asymptotic
quasinormal mode spectrum of non-rotating black holes'',
arXiv:gr-qc/0307020 (2003).
%%CITATION = GR-QC 0307020;%%


\bibitem{X1} K.H.C. Castello-Branco, private communication (2003).

\bibitem{X2} R.A. Konoplya, private communication (2003).





%\bibitem{Kon1} R.A. Konoplya, ``On quasinormal modes of
%small Schwarzschild-anti de Sitter black holes'', Phys. Rev. D {\bf 66},
%044009 (2002) [hep-th/0205142].
%%%CITATION = HEP-TH 0205142;%%





\end{thebibliography}
\end{document}